\documentclass[letter,twocolumn]{jpsj3} 

\usepackage{graphicx}
\usepackage{amssymb}
\usepackage{bm}
\usepackage{color}
\usepackage{ulem}

\newcommand{\red}[1]{\textcolor{black}{#1}}
\newcommand{\blue}[1]{\textcolor{black}{#1}}

\title{Electronic Structure \blue{Calculation and Superconductivity} in $\lambda$-(BETS)$_{2}$GaCl$_{4}$}

\author{
Hirohito Aizawa$^{1}$\thanks{aizawa@kanagawa-u.ac.jp},
Takashi Koretsune$^{2, 3}$,
Kazuhiko Kuroki$^{4}$, and 
Hitoshi Seo$^{5, 6}$
}

\inst{
$^1$Institute of Physics, Kanagawa University, Yokohama, Kanagawa 221-8686, Japan\\
$^2$Department of Physics, Tohoku University, Sendai 980-8578, Japan\\
$^3$JST PRESTO, Kawaguchi, Saitama 332-0012, Japan\\
$^4$Department of Physics, Osaka University, Toyonaka, Osaka 560-8531, Japan\\
$^5$Condensed Matter Theory Laboratory, RIKEN, Wako, Saitama 351-0198, Japan\\
$^6$Center for Emergent Matter Science (CEMS), RIKEN, Wako, Saitama 351-0198, Japan
}

\abst{
Quasi-two-dimensional molecular conductor $\lambda$-(BETS)$_2$GaCl$_4$ shows 
superconductivity (SC) below 5.5~K, neighboring the dimer-type Mott insulating phase. 
To elucidate the origin of SC and its \blue{gap function}, 
we carry out first-principles band calculation and derive a four-band tight-binding model 
from the maximally localized Wannier orbitals. 
Considering the spin-fluctuation-mediated mechanism 
by adding the Hubbard $U$-term to the model, 
we analyze the SC gap function by applying the random phase approximation. 
We show that the SC gap changes its sign four times 
along the Fermi surface (FS) in the unfolded Brillouin zone, 
suggestive of a \blue{$d$-wave-like SC gap, }
\blue{which only has two-fold symmetry because of} 
the low symmetry of the crystal structure. 
\blue{Decomposing the SC 
gap into the pairing functions along the crystal axes, 
we compare the result to similar analysis of} 
the well-studied $\kappa$-type molecular conductors 
and to the experiments. 
}

\begin{document}
\maketitle

The quasi-two-dimensional (Q2D) molecular conductor $\lambda$-(BETS)$_{2}$GaCl$_{4}$, 
where BETS is bis(ethylenedithio)tetraselenafulvalene, 
exhibits superconductivity (SC) below 5.5~K~\cite{Kobayashi1993a,Kobayashi1993,Kobayashi2004}. 
It has attracted interest as a candidate for realizing the FFLO state 
under 
magnetic field owing to its highly two-dimensional electronic structure~\cite{Tanatar2002,Coniglio2011,Uji2015}. 
Another interest is that its isostructural compound 
$\lambda$-(BETS)$_{2}$FeCl$_{4}$ shows a field-induced SC phase 
under 
strong magnetic field,~\cite{Uji2001,Balicas2001}
considered to be connected to that of the Ga salt 
(at zero field) 
as indicated by the measurements of alloyed samples~\cite{Uji2003}.
Despite extensive experimental works, 
theoretical investigation of SC in this compound from a microscopic viewpoint has been lacking, 
which is the purpose of this study.

In the $\lambda$-type structure, the BETS molecules stack along the $a$ direction, 
forming a triclinic unit cell with space group P$\bar{1}$~\cite{Kobayashi1993a,Kobayashi1993}.
There are dimers of BETS molecules with large intradimer transfer integrals (termed $t_{\rm A}$), 
which show further dimerization, i.e., a tetramer of BETS forms the unit cell.
The GaCl$_{4}^{-1}$ closed-shell anion sheets lead to the highest-occupied molecular orbital (HOMO) of BETS forming a Q2D quarter-filled system in terms of holes.~\cite{Kobayashi2004}
From its dimerized structure, whose limit of large dimerization will be a half-filled system, the electronic structure has an analogy with the well-studied Mott transition system $\kappa$-(ET)$_2X$ [ET = bis(ethylenedithio)tetrathiafulvalene].
In fact, by chemical substitution in the anions Ga$X_{z}Y_{4-z}$ ($X, Y$=F, Cl, Br)~\cite{Kobayashi1997,Tanaka1999} or by choosing different donor molecules~\cite{Mori2001}, 
the SC phase is suggested to locate next to the Mott insulating phase as in $\kappa$-(ET)$_{2}X$~\cite{Kanoda2006}.

Although the nature of the insulating state just near the SC phase remains to be clarified, i.e., whether it is non-magnetic~\cite{Kobayashi1997,Seo1997} or antiferromagnetic~\cite{Mori2001,Saito2018}, 
a recent NMR measurement in $\lambda$-(BETS)$_{2}$GaCl$_{4}$ reports the development of spin fluctuations above the SC transition temperature~\cite{Kobayashi2017}.
As for 
\blue{the SC gap function}, early measurements show a two-fold symmetry within the conductive plane, by means of the anisotropy of the upper critical field $H_{\rm c2}$~\cite{Tanatar1999} and of the flux-flow resistivity~\cite{Yasuzuka2014}.
\blue{
Ref.~\citen{Yasuzuka2014} 
observes a dip structure in the angle dependence of the resistivity under magnetic field, 
when the magnetic field is applied parallel to the $c$ axis. 
}
More recently, a heat capacity measurement indicated 
the line-nodal gap of $d$-wave pairing~\cite{Imajo2016}, 
whereas a $\mu$SR measurement reports a possible mixture of the extended $s$- and $d$-wave 
\blue{gaps}~\cite{Dita2018-thesis}.

The electronic structure of $\lambda$-(BETS)$_{2}$GaCl$_{4}$ has been discussed within the tight-binding model based on the HOMO of the BETS molecule, 
where the transfer integrals are calculated using the extended H\"{u}ckel method~\cite{Tanaka1999,Kobayashi1996,Mori2002}.
The band structure near the Fermi energy shows four bands since the unit cell contains four BETS molecules as mentioned above. 
The calculated Fermi surface (FS) is similar to that of the $\kappa$-(ET)$_{2}X$,~\cite{Oshima1988} 
which consists of a pair of open and closed FS, despite the difference in their molecular packings. 
One issue is that, since the extended H\"{u}ckel method contains semi-empirical parameters, 
there are estimates with appreciable discrepancies. 

In this study, we present the band structure obtained from first-principles calculations, 
and estimate the transfer integrals of the four-band model from the maximally localized Wannier orbitals (MLWO). 
Then, considering the pairing mechanism mediated by the spin fluctuations, 
we apply the random phase approximation (RPA) to the four-band Hubbard model of $\lambda$-(BETS)$_{2}$GaCl$_{4}$. 
The results show a $d$-wave-like SC gap and we will discuss its origin related to the spin susceptibility.

\begin{figure}[!htb]
\centering
\includegraphics[width=8.0cm]{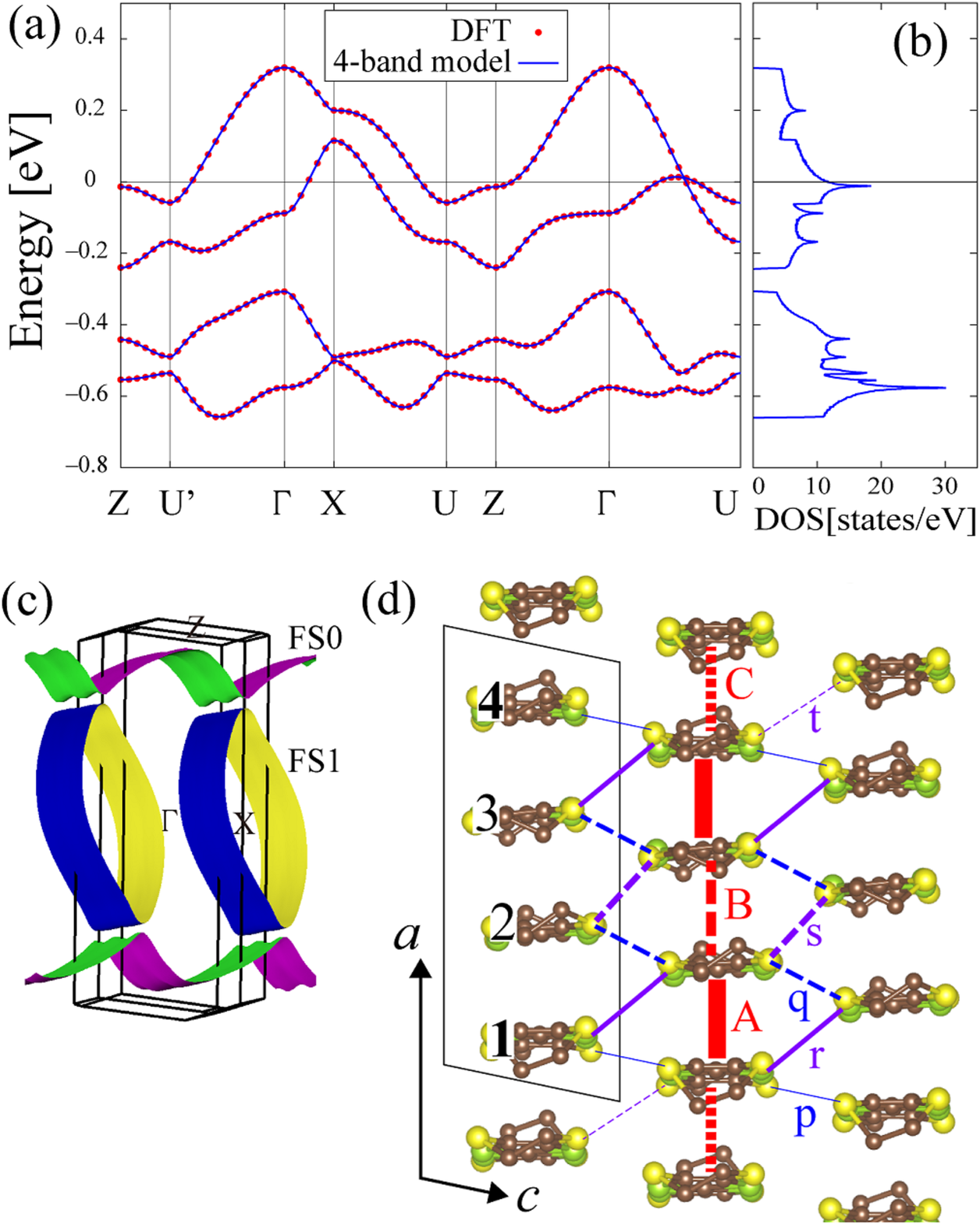}
\caption{
(a) Band structure obtained from 
the DFT calculation (red dotted curves) 
and the four-band model (blue solid curves), 
where the Fermi level is taken as zero energy. 
X, Z, U and U' represent 
(1/2, 0, 0), (0, 0, 1/2), (1/2, 0, 1/2) and ($-$1/2, 0, 1/2), respectively. 
(b) Density of states (DOS) obtained from the four-band model. 
(c) The DFT Fermi surface (FS) 
where 
FS0 (FS1) corresponds to the open (closed) FS (see text). 
(d) The transfer integrals in the four-band model, 
where each BETS molecule is considered as a lattice site; 
the parallelogram represents the unit cell. 
The notation of the transfer integrals is the same as 
in the previous studies~\cite{Kobayashi1996,Mori2002}. 
The BETS molecules numbered in the bold (normal) characters are related through inversion symmetry. 
}
\label{fig1}
\end{figure}

The first-principles band calculations were performed within density functional theory (DFT) 
with generalized gradient approximation~\cite{Perdew1996} 
using WIEN2k~\cite{WIEN2K}, 
and a tight-binding model was derived 
by applying MLWO~\cite{Marzari1997,Kunes2010} 
scheme using wannier90 package~\cite{wannier90-Mostofi2014}. 
Figure~\ref{fig1}(a) shows the two-dimensional band dispersions near the Fermi level for the experimental structure data~\cite{Tanaka1999}. 
Dispersion along the interlayer direction is small, of the order of 0.1~meV. 
There are four bands originated from HOMO of BETS, corresponding to the extended H\"{u}ckel bands. 
One point we note is that, since nearly flat band dispersions are present near Z point, 
the density of states (DOS) exhibits 
a van-Hove singularity \blue{(vHS)} slightly below the Fermi level, 
as shown in Fig.~\ref{fig1}(b). 
In Fig.~\ref{fig1}(c), we show the FS obtained from the DFT calculation, 
consisting of open and closed portions, 
which we call FS0 and FS1 in the following. 
The former comes from the top band and the latter comes from the second-to-top band.
The shape of the FS is similar to the extended H\"{u}ckel results~\cite{Kobayashi1996,Tanaka1999,Mori2002}. 
We regard these four bands as the target bands and derive a tight-binding model by constructing a MLWO on each molecule. 
As shown in Fig.~\ref{fig1}(a), the band structure of the four-band model, which includes the distant transfer integrals, accurately reproduces the DFT band dispersion.

%
%

%
\begin{table}[!htb]
\centering
\caption{
Transfer integrals and site-energy difference in meV 
for $\lambda$-(BETS)$_2$GaCl$_4$, 
where the site-energy difference between the BETS-1(4) and BETS-2(3) 
is defined as $\Delta E \equiv \ E_{2(3)}-E_{1(4)}$. 
The superscript, eH, stands for the extended H\"{u}ckel results~\cite{Kobayashi1996,Tanaka1999,Mori2002} 
and the subscript, Fe, represents the 
results of $\lambda$-(BETS)$_{2}$FeCl$_{4}$, 
having the same crystal structure. 
}
\begin{tabular}{ l|r|r|r|r } \hline
Label & $t$ & 
$t^{\rm eH}$~\ \cite{Tanaka1999} 
& $t_{\rm Fe}^{\rm eH}$~\ \cite{Kobayashi1996} 
& $t_{\rm Fe}^{\rm eH}$~\ \cite{Mori2002} \\ \hline \hline
A  &    233  &    238   &    747   &    336 \\ 
B  & $-$131  &  $-$98   & $-$302   & $-$183 \\
C  & $-$138  &  $-$58   & $-$279   & $-$148 \\
p  &     15  &     13   &      3   &     28 \\
q  &     59  &     31   &    189   &     93 \\
r  &     63  &     37   &    237   &    130 \\
s  &  $-$82  &  $-$48   & $-$176   & $-$171 \\
t  &  $-$17  &   $-$4   &  $-$33   &  $-$26 \\ \hline
$\Delta E$
   &  $-$29  &     --   &     --   &     -- \\ \hline
\end{tabular}
\label{table-4band}
\end{table}
%

%
We summarize the obtained transfer integrals 
and the energy difference between the nonequivalent BETS, $\Delta E$, 
in Table~\ref{table-4band}, together with the extended H\"{u}ckel results in the literature. 
The notation of inter-molecular bonds is as shown in Fig.~\ref{fig1}(d). 
The other transfer integrals not listed here 
have absolute values less than 13~meV. 
In the RPA analysis below, 
we use all the obtained transfer integrals 
in the two-dimensional plane.~\cite{distant-tij}
As a common feature among our results and the previous extended H\"{u}ckel calculations, 
$t_{\rm A}$ is the largest, which is the intradimer transfer integral.
This gives the splitting between the upper two and lower two bands, 
approximately corresponding to the antibonding and bonding combinations of the HOMO. 
The transfer integrals along the stacking direction 
$t_{\rm B}$ and $t_{\rm C}$ have close values 
in contrast with previous data, 
which indicates that the degree of tetramerization is smaller than previously discussed.~\cite{Seo1997} 

The effective transfer integrals between the anti-bonding combination of HOMO of BETS dimers 
along the $a$ direction can be approximated from the large dimerization limit 
as $\tilde{t}_{\rm B} \equiv t_{\rm B}/2$ and $\tilde{t}_{\rm C} \equiv t_{\rm C}/2$; 
that in the transverse direction is $t_\perp \equiv \ \left( t_{\rm p} +  t_{\rm q} + t_{\rm r} \right) /2$. 
Our results show 
a relation $|\tilde{t}_{\rm B}| \simeq |\tilde{t}_{\rm C}| \simeq t_\perp$. 
Then, the BETS dimers possess a square-lattice-like network along the $a$ and $c$ directions, 
with weaker diagonal transfer integrals 
$\tilde{t}_{\rm s} \equiv t_{\rm s}/2 \simeq 0.6 t_\perp$ 
or $\tilde{t}_{\rm t}  \equiv t_{\rm t}/2 \simeq 0.1 t_\perp$ 
along the $a+c$ direction. 
We can interpret the large DOS to be originated from this relation 
since the ideal square lattice has a singularity of the DOS at half-filling. 
Another recent DFT calculation based on the pseudopotential shows the same result~\cite{Dita2018-thesis}.

Next, by introducing the on-site Coulomb interaction $U$ to the four-band model, 
we study the spin susceptibility $\chi_{\rm sp}$ and the SC gap \blue{function}  
within the framework of the spin-fluctuation-mediated pairing mechanism.
The Hamiltonian is described as 
\begin{eqnarray}
 H&=&\sum_{\left< i \alpha: j \beta \right>, \sigma}
  \left\{ t_{i \alpha: j \beta} 
   c_{i \alpha \sigma}^{\dagger} c_{j \beta \sigma} + {\rm H. c.} 
  \right\} 
\nonumber \\
  &+&
  \Delta E \sum_{i, \alpha=2, 3} n_{i \alpha}
  +\sum_{i, \alpha} U
  n_{i \alpha \uparrow} n_{i \alpha \downarrow}, 
  \label{Hij}
\end{eqnarray} 
where $i$ and $j$ are unit-cell indices, 
$\alpha$ and $\beta$ specify the sites 1--4 in a unit cell [see Fig.~\ref{fig1} (d)], 
$c_{i \alpha \sigma}^{\dagger}$ ($c_{i \alpha \sigma}$) is the creation (annihilation) operator for spin $\sigma$ at site $\alpha$ in unit cell $i$. 
$t_{i \alpha: j \beta}$ is the transfer integral 
between site $(i, \alpha)$ and site $(j, \beta)$, 
estimated as above, 
and $\left< i \alpha: j \beta \right>$ represents the site pairs. 
$n_{i \alpha \sigma}$ is the number operator for electrons with spin $\sigma$ on site $\alpha$ in unit cell $i$ 
and $n_{i \alpha}=n_{i \alpha \uparrow}+n_{i \alpha \downarrow}$. 

To deal with the effect of the Coulomb interaction $U$, we apply 
the multisite RPA, e.g., 
described in Ref.~\citen{A_Kobayashi2004}; 
here we focus on SC state in a situation where other instabilities are weaker. 
The Green's function, as well as the susceptibilities, pairing interaction, and SC gap function are all 4$\times$4 matrices. 
The gap function $\hat{\varphi}\left( \textbf{\textit{k}}, i \varepsilon_{n} \right)$ and its eigenvalue $\lambda$ are obtained by solving the linearized Eliashberg equation. 
The critical temperature $T_{\rm c}$ corresponds to the temperature where $\lambda$ reaches unity. 
Because we consider only the on-site interaction $U$, the spin susceptibility is much larger than the charge susceptibility. 
Therefore, we will show the spin susceptibility $\chi_{\rm sp}$ obtained from the largest eigenvalue for 
\blue{the lowest Matsubara frequency.} 
The SC gap function is presented in the band representation 
at the lowest Matsubara frequency. 
In the present calculation, we take 96$\times$96 $k$-point meshes 
and 16384 Matsubara frequencies.
The on-site interaction is chosen as $U=0.4$~eV.

\begin{figure}[!htb]
\centering
\includegraphics[width=8.0cm]{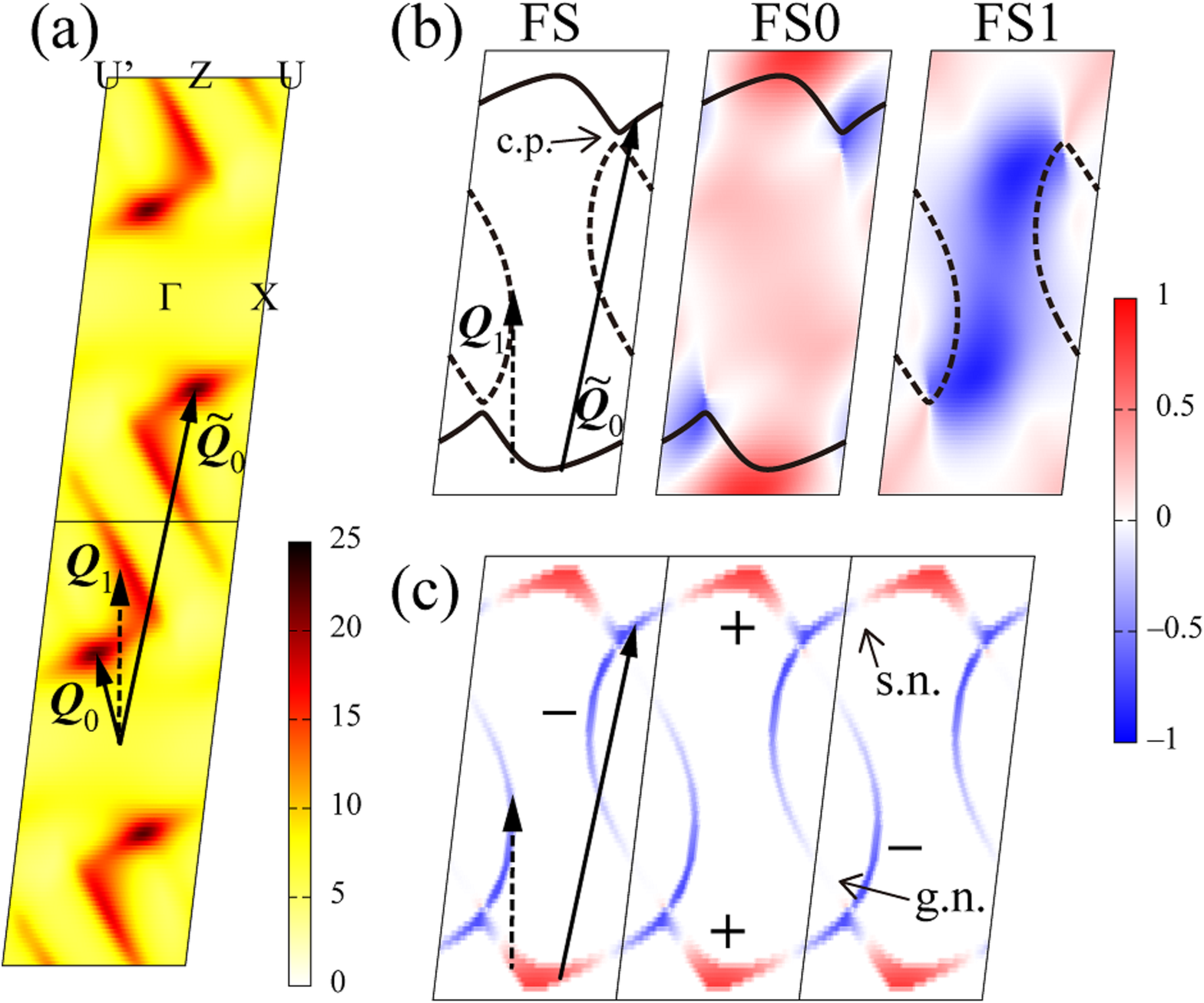}
  \caption{
(a) Spin susceptibility
\blue{, whose unit is 1/eV, }
at $T$=0.006~eV, where the black solid (dashed) arrow represents 
the nesting vector $\mbox{\boldmath $Q$}_0$ ($\mbox{\boldmath $Q$}_1$) and  
the arrow $\boldsymbol{\tilde{Q} }_0 = -\boldsymbol{Q}_0 + \left(0, \red{2} \pi \right)$. 
(b) The left panel is the FS and its nesting vectors, 
where ``c.p." stands for the crossing point (see text) 
\blue{and the solid (dashed) curves represent the FS0 (FS1).} 
The center (right) panel shows the SC gap for FS0 (FS1), 
where the red (blue) contours represent the positive (negative) 
SC gap sign\blue{, where the plotted SC gap represents the ratio to its maximum value}. 
In the center and right panels, the thick 
\blue{black} 
curves represent the FS for which the gap function is plotted\blue{.} 
(c) The SC gap function within $\pm 0.01$~eV from 
the Fermi level, where \blue{``g.n. (s.n.)"} stands for \blue{``gentle (steep) node" structure (see text).}
}
\label{fig2}
\end{figure}

As shown in Fig.~\ref{fig2}(a), at temperature $T=0.006$~eV ($\simeq$ 70 K), 
the spin susceptibility 
$\chi_{\rm sp}$ has the maximum value around $\mbox{\boldmath $Q$}_0 = \left( Q_{0a}, Q_{0c} \right) \simeq \left( -3\pi/8, 3\pi/8 \right)$ 
and a broad substructure around $\mbox{\boldmath $Q$}_1 = \left( Q_{1a}, Q_{1c} \right) \simeq \left( -\pi/6, 5\pi/6 \right)$.
To discuss the nesting properties in the following, 
$\chi_{\rm sp}$ in the extended zone along the $\Gamma$-Z direction is shown and we define 
$\boldsymbol{\tilde{Q} }_0 =-\boldsymbol{Q}_0 + \left(0, \red{2}\pi \right)$. 
In Fig.~\ref{fig2}(b), we show the FS on the left and the SC gap function 
for the top (second-to-top) band, namely, for FS0 (FS1) on the center (right) 
for $\lambda \simeq 0.42$~\blue{\cite{eigenvalue-lambda}}. 
The FS can approximately be regarded as an ellipse in the unfolded Brillouin zone, 
whereas FS0 and FS1 are slightly disconnected around the $k$ points 
where they approach to each other. 
In the following, we will call this point as the crossing point and 
the elliptic FS in the unfolded zone as the `extended FS'. 
As we can see in the figure, 
wave-number vectors $\boldsymbol{\tilde{Q} }_0$ and $\mbox{\boldmath $Q$}_1$ 
correspond to the FS nesting. 
As for the SC gap function, 
first we note that for FS0 (FS1) it has a positive (negative) sign 
along almost the whole Brillouin zone. 
This gives rise to the large $s$-wave components in the analysis below.

%
To clarify the relation between the electronic structure 
and the SC gap, 
in Fig.~\ref{fig2}(c) we plot the SC gap functions 
for {\boldmath $k$} vectors within $\pm 0.01$~eV from the Fermi level, 
which corresponds to be about 1\% of the band width (of four bands), 
from the Fermi level in the extended zone.
Then, we can see that the SC gap changes its sign four times 
along the extended FS reminiscent of a 
\blue{$d$-wave-like gap, which possesses two kinds of nodes, 
from which gentle/steep increase of the SC gap is seen. 
We call them as gentle/steep node structures.} 
In the figure, the two nesting vectors $\boldsymbol{\tilde{Q} }_0$ 
and $\mbox{\boldmath $Q$}_1$ connect the portions of FS 
where the SC gap has a large amplitude and shows sign changes. 
The positions of the FS where the SC gap amplitude is large 
almost coincide with the positions giving rise to the 
\blue{vHSs}, 
resulting in a high stability of the gap structure. 
In fact, a similar analysis based on the effective two-band model (the upper two bands) 
also gives rise to similar angular dependence.~\cite{Dita2018-thesis}

Next, we attempt to decompose the SC gap into different components, as has been done for 
$\kappa$-(ET)$_{2}X$~\cite{Powell2007,Guterding2016a,Guterding2016,Watanabe2017}. 
In the $\lambda$-type structure, 
the point group symmetry is low as $C_i$, 
therefore naturally different components mix.~\cite{Powell2006}
\blue{Therefore, we decompose the $d$-wave-like gap into pairing components along the crystal axes.} 
Here we take \blue{crystal $c$} and \blue{$a$} directions as $x$ and $y$ axes, 
respectively, to make the correspondence between other systems clearer. 
In this choice of axes, 
the SC gap structure in Fig.~\ref{fig2} (c) apparently looks close to 
a $d_{x^2-y^2}$-wave \blue{gap} 
since the nodes are nearly along the diagonal directions. 
We introduce the fitting function given as 
\blue{
\begin{eqnarray}
\varphi^{f}(\mbox{\boldmath $k$}) &=& 
  C_{0}^{f} 
+ C_{c}^{f} \cos(k_x)
+ C_{a}^{f} \cos(k_y)
\nonumber \\
& &
+ C_{c+a}^{f} \cos(k_x + k_y)
+ C_{c-a}^{f} \cos(k_x - k_y)
\nonumber \\
& &
+ \cdots
\textrm{(up to 20th nearest neighbors)}, 
\label{fitting-function}
\end{eqnarray}
}
where $f$ is for the choice of the two bands represented by FS0 or FS1, 
the subscript \blue{represents the pairing direction,} 
and the fitting variables \blue{$C^{f}$} are 
the weights of the basis function on the FS ``$f$". 
Longer range pairing states \blue{as represented in Eq.~(\ref{fitting-function})} 
are also considered in the actual calculation, but have small contributions. 
We summarize the ratio of the fitting variables 
for the basis function in Table~\ref{table2}. 
In the case of FS0 [center of Fig.~\ref{fig2} (b)], 
\blue{although the ratio in the $c-a$ direction is the largest, 
the pairing ratio along the $a$ direction is subdominant 
and comparable with that of the intra-unit cell. 
It is suggestive that the SC gap of the FS0 is affected by 
the pairing along the $a$ direction, 
in which the BETS molecules stack. 
By contrast, for FS1, the three components, 
namely the intra-unit cell as well as $c$ and $a$ directions, are comparable. 
As expected, the SC gap of FS1 exhibits a two-dimensional pairing. 
}

\begin{table}[!htb]
\centering
\caption{
Ratio of the fitting variables of the basis function on the FS ``$f$", 
from the \blue{$d$-wave-like gap function.}
We take the \blue{intra-unit-cell} component \blue{$C_0^f$} as unity; 
to stress the different sign between the two bands, 
we put different signs. 
}
\blue{
\begin{tabular}{l|r|r } \hline
\hspace{0pt} Fitting variable \hspace{0pt} & 
\hspace{1pt} FS0     \hspace{1pt} & 
\hspace{0pt} FS1     \hspace{0pt} \\ \hline \hline
$C^{f}_0$     &    1.00  & $-$1.00   \\ 
$C^{f}c$      & $-$0.22  & $-$1.00   \\
$C^{f}a$      &    1.27  & $-$1.38   \\
$C^{f}{c+a}$  &    0.14  & $-$0.09   \\
$C^{f}{c-a}$  & $-$1.40  & $-$0.41   \\ \hline
\end{tabular}
}
\label{table2}
\end{table}

\blue{
To compare with the previous studies of 
$\kappa$-(ET)$_{2}X$~\cite{Powell2007,Guterding2016a,Guterding2016,Watanabe2017}, 
we rewrite the ratio of the well-known SC gap, 
as $d_{x^2-y^2}$-, $d_{xy}$-, extended $s_{1(2)}$-wave, 
which is the pairing with the same sign between the first (second) nearest neighbors. 
Note that we decompose the $d$-wave-like gap into the well-known SC gap 
and confirm that the same components are obtained.  
We list the components of the SC gap in Table~\ref{table3}. 
Several SC-gap components of the FS0 are comparable.} 
By contrast, for the FS1, the components of the isotropic $s$- and 
extended $s_1$-wave \blue{possess large negative value.}
We should note that, 
even though the component of the isotropic 
$s$-wave\blue{, which is same as the intra-unit-cell pairing,} 
is large, 
this does not mean that the pairing, in real space picture, 
occurs on the same BETS molecule, 
since the SC components are obtained in the ``folded" Brillouin zone. 
Namely, an anisotropic pairing, e.g., 
the nearest neighbor pairing between BETS-2 and BETS-1 
or BETS-3 within the same unit cell, 
is converted to an isotropic $s$-wave component 
in the folded Brillouin zone 
because the pairing occurs within the unit cell.

\begin{table}[!htb]
\centering
\caption{
Ratio of the \blue{component of the well-known SC gap on the both FSs based on Table~\ref{table2}.} 
}
\begin{tabular}{ l|r|r } \hline
\hspace{1pt} \blue{SC gap component} \hspace{1pt} & 
\hspace{1pt} FS0           \hspace{1pt} & 
\hspace{0pt} FS1           \hspace{0pt} \\ \hline \hline
   \blue{${\rm isotropic}~s$-wave} &    1.00  & $-$1.00   \\ 
\blue{${\rm extended}~s_{1}$-wave} &    0.53  & $-$1.19   \\
                $d_{x^2-y^2}$\blue{-wave} & $-$0.75  &    0.19   \\
\blue{${\rm extended}~s_{2}$-wave} & $-$0.63  & $-$0.25   \\
                     $d_{xy}$\blue{-wave} &    0.77  &    0.16   \\ \hline
\end{tabular}
\label{table3}
\end{table}
%

%
The results here that multiple components have comparable values are noticeably different from the case of $\kappa$-(ET)$_2X$.~\cite{Powell2007,Guterding2016a,Guterding2016,Watanabe2017} 
In that case, the effective half-filled dimer Hubbard model 
shows the instability toward $d_{x^{2}-y^{2}}$-wave SC in the extended zone~\cite{Kino1998,Kondo1998,Schmalian1998,Kuroki2002,Kyung2006,Powell2007,Zantout2018} 
while for the 3/4-filled model realistic parameters provide 
$d_{xy}$-type 
\blue{-wave gap}~\cite{Kuroki2002,Silva2016,Guterding2016a,Guterding2016,Watanabe2017,Zantout2018} 
but with considerable 
extended $s$-\blue{wave} 
component.~\cite{Powell2007,Guterding2016a,Guterding2016,Watanabe2017} 
We can attribute such difference to the different crystal structure geometries: 
$\kappa$-type has a $D_{2h}$ point group symmetry, 
so that pure $d_{x^{2}-y^{2}}$-wave can be stabilized 
but not pure $d_{xy}$\blue{-wave} in the extended zone. 
$\kappa$-(ET)$_2X$ has parameters close to 
the triangular lattice 
giving rise to geometrical frustration effect, 
while our analysis here provides a square-lattice like network, 
as in the high $T_{\rm c}$ cuprates producing 
the stability of $d_{x^{2}-y^{2}}$-wave, 
but with large mixing with other components 
\blue{of the well-known SC gap} 
due to the low symmetry of the crystal structure. 

Finally let us discuss the experimental works 
from the viewpoint of our \blue{results giving the $d$-wave-like gap}. 
The results in the transport measurements indicating 
the two-fold symmetry of the angular dependence of 
\blue{the SC gap in this compound} 
are compatible with our results since 
\blue{the $d$-wave-like gap}
only possesses 
the two-fold symmetry~\cite{Tanatar1999,Yasuzuka2014}. 
The existence of the nodal SC gap is suggested by 
a recent measurement of the heat capacity~\cite{Imajo2016}, 
which is in accordance with our results showing nodes along the diagonal directions. 
\blue{As for the nodal position, the flux-flow resistivity measurement suggests 
a dip structure of the resistivity when the magnetic field is applied 
parallel to the $c$ axis~\cite{Yasuzuka2014}. 
This is consistent with the $d$-wave-like gap we obtained, 
namely, the large SC gap around vHS and the steep node structure are present. 
}
A recent $\mu$SR measurement suggests 
that the SC of this compound is a mixture of 
the extended $s$-wave and $d$-wave SC~\cite{Dita2018-thesis}. 
A direct comparison between our results might be difficult 
since the method of decomposing the \blue{SC gap} is different from ours here, 
magebut nevertheless, the mixture of different components is indeed consistent.

In conclusion, we have obtained the DFT band structure 
and the four-band model of the Q2D molecular conductor $\lambda$-(BETS)$_{2}$GaCl$_{4}$.
Within the spin-fluctuation-mediated pairing mechanism, 
we study the SC gap function and its properties by applying the RPA. 
The network of the BETS dimers shows a square-lattice-like structure, 
giving rise to large DOS near the Fermi level. 
We propose that the FS nesting within 
this characteristic electronic structure results in the 
\blue{$d$-wave-like} SC gap, 
which changes its sign four times 
along the extended FS 
\blue{and possesses the two-fold symmetry.}

To elucidate the pairing components \blue{of the $d$-wave-like gap,  
we have decomposed this gap function into the pairing components 
along the crystal axes, 
and estimate the pairing ratio for each FS. 
We have shown that the SC gap of FS0 is affected 
by the pairing in the stacking direction of the BETS 
and that the gap of FS1 exhibits a two-dimensional pairing. 
To compare the previous studies of $\kappa$-(ET)$_2X$, 
we transform the component of the pairing in the crystal axes 
to that of the well-known SC gap functions, and show that 
the several SC gap components can be comparable in both FSs. 
}

The effect of strong electronic correlation beyond RPA, 
which is expected to play a role since the system 
is considered to be located near the Mott transition, 
is an interesting issue left for future studies.

\section*{Acknowledgments}

The authors acknowledge D. P. Sari and I. Watanabe for valuable discussions. 
HA is grateful to S. Yasuzuka and S. Imajo for useful discussions. 
This work is supported by 
the Japan Society for the Promotion of Science KAKENHI 
Grants No. 16K17754, 18K03442 and 26400377, 
Grants-in-Aid from the Yokohama Academic Foundation, 
and the RIKEN iTHES Project.


\begin{thebibliography}{99}
\bibitem{distant-tij}
The transfer integrals are included up to the third (seventh) nearest neighbor unit cell along the $a$ ($c$) direction. 
Note that the four BETS molecules stack along the $a$ direction in the unit cell as shown in Fig.~\ref{fig1}~(d).
\bibitem{eigenvalue-lambda}
\blue{
Larger values of $\lambda$ can be obtained upon further lowering the temperature. 
However, at such low temperatures, the spin susceptibility exhibits an extremely sharp structure around the nesting vector. 
In fact, this is an artifact of the present RPA approach, which does not take into account the self-energy effect. 
Such a spin structure results in an unphysical $\mbox{\boldmath $k$}$-space variance of the superconducting gap. 
Hence, here we perform the calculation at sufficiently high temperature where the gap function remains physically sound, although $\lambda$ is somewhat small. 
}
\end{thebibliography}
\end{document}